\documentclass[epj,nopacs]{svjour}
\usepackage{amsmath}
\usepackage{titlesec}
\usepackage[titletoc,title]{appendix}

\usepackage{pdfpages}
\usepackage{vmargin} 
\setmargins{2.5cm}{1.5cm}{16.5cm}{23.42cm}{10pt}{1cm}{0pt}{2cm}
\usepackage{appendix}
\renewcommand{\tablename}{Tabla}
\usepackage[english]{babel}
\usepackage[utf8]{inputenc}
\usepackage{csquotes}
\usepackage{braket}
\usepackage{graphicx} 
\usepackage{amsmath} 
\usepackage{amssymb} 
\usepackage[utf8]{inputenc}
\numberwithin{equation}{section}
\usepackage[utf8]{inputenc}
\usepackage[colorlinks=true, allcolors=black]{hyperref}
\usepackage {amsmath}
\usepackage[maxbibnames=3,
backend=biber,
style=ieee,
isbn=false,
language=english
]{biblatex} 
\usepackage [pdftex]{}

\begin{document}
\title{The Wigner formalism on black hole geometries
}
\author{Jose A.R. Cembranos\inst{1,2}
\thanks{cembra@ucm.es}
 \and 
David García García \inst{1}\thanks{davgar27@ucm.es}
}                     
%
%
\institute{Departamento de Física Teórica, Universidad Complutense de Madrid, Plaza de Ciencias 1, Facultad de Ciencias Físicas, 28040, Madrid, Spain. \and Institute of Particle and Cosmos Physics (IPARCOS), Universidad Complutense de Madrid, Plaza de Ciencias 1, Facultad de Ciencias Físicas, 28040, Madrid, Spain.}
%
\abstract{
This work explores the intersection of quantum mechanics and curved spacetime by employing the Wigner formalism to investigate quantum systems in the vicinity of black holes. Specifically, we study the quantum dynamics of a probe particle bound to a Schwarzschild black hole using a phase-space representation of quantum mechanics. The analysis begins with a review of the covariant Wigner function in curved spacetime, highlighting its application to spherically symmetric, uncharged black holes. We then derive an effective potential from the Schwarzschild metric, which defines the Hamiltonian for the electron. Relativistic corrections are treated perturbatively to estimate energy levels and associated Wigner functions for the bound state. Additionally, we compare the results obtained through the Schrödinger equation with those derived directly using the symplectic formalism, demonstrating the consistency and versatility of the phase-space approach. The study sheds light on quantum behavior near black holes and suggests new avenues for combining quantum kinetic theory with relativistic gravitational settings.}

\maketitle
\pagestyle{plain}
\newpage

\section{Introduction}\label{sec:intro}

The exploration of quantum mechanics in curved spacetime is a rich and evolving field, intersecting with foundational concepts in general relativity and quantum field theory. Among the notable efforts is the quest to define a covariant Wigner function—a phase-space representation of quantum mechanics that maintains covariance in curved spacetime geometries. Several attempts have been made in this direction \parencite{HABIB1989335,PhysRevD.37.2165,PhysRevD.56.920,PhysRevD.37.2901}. Notably, \textcite{Emelyanov_2017} proposed a covariant Wigner function for the Schwarzschild metric, which describes spherically symmetric, uncharged black holes. In their work, they used quantum kinetic theory to investigate the behavior of a scalar field conformally coupled to gravity near and far from the black hole.

To achieve this, the authors employed the conformal symmetry of the scalar field to derive the Wightman function and subsequently used it to construct the Wigner function in different spacetime regions. Their analysis revealed interesting, albeit problematic, physical interpretations: while macroscopic variables derived from the Wigner function matched expectations far from the black hole, closer to the horizon, negative particle densities and imaginary entropy densities emerged. They attributed these anomalies to quantum fluctuations potentially induced by the black hole's evaporation, drawing parallels with the Casimir effect.

Complementing this perspective, the phase-space formalism of quantum mechanics offers a powerful and alternative approach to understanding quantum systems in curved spacetime. Originating in the early 20th century with contributions from Weyl, Wigner, Moyal, and Groenewold, symplectic quantum mechanics treats positions and momenta on equal footing and is mathematically equivalent to the Hilbert space formalism \parencite{WEYL31,Wigner32,Moyal49,Groenewold1946}. Recent developments have further enriched this approach, such as the application of the Wigner formalism to the Aharonov–Bohm effect \parencite{Cembranos2024}, the use of the Segal–Bargmann formalism to study Wehrl entropy in entangled systems \parencite{AlonsoLopez2023}, and the formulation of functional quantum field theory in phase space \parencite{Cembranos2021}. These contributions highlight the flexibility of phase-space methods in capturing quantum phenomena across diverse contexts, including systems with non-trivial geometries and interactions.

In this work, we bridge these two domains by employing the Wigner formalism to study the quantum dynamics of a bound state consisting of an electron and a black hole. Using the Schwarzschild metric, we derive an effective potential analogous to the Newtonian case, which defines the Hamiltonian of the system. Treating the relativistic corrections as perturbations, we calculate the energy levels of the electron and obtain the associated Wigner functions. Additionally, we compare these results with solutions obtained through the symplectic formalism, bypassing the Schrödinger equation entirely.

The unification of phase-space quantum mechanics with curved spacetime analysis not only deepens our understanding of quantum systems near black holes but also provides new tools to address long-standing questions about the quantum properties of spacetime itself. This paper lays the groundwork for further exploration, particularly in extending the Wigner formalism to more complex spacetimes and fields.

\vspace{1cm}
\section{Phase-space formulation of quantum mechanics: Wigner-Weyl formulation} \label{sec:Formalism}
Quantum mechanics is a non-commutative theory, so it is necessary to define a relationship between operators and variables, which can be done in various ways. However, the expectation
value of an arbitrary operator that represents a certain quantum-mechanical observable can be calculated via a distribution function F(q,p,t) as:
    \begin{equation}
        Tr\left[\hat{\rho}(\hat{q},\hat{p},t)\hat{A}(\hat{q},\hat{p})\right]=\iint dqdp\ A(q,p,t)F(q,p,t),
        \label{eq:FdistrPS}
    \end{equation}
where $A$ is a scalar function that is obtained by replacing the operators $\hat{p}$ and $\hat{q}$ for scalar variables p y q in the expression for $\hat{A}$ using a determined rule of association. Nevertheless, a unique way to define the function $F$ does not exist because of the non-commutativity of the operators. In order, to understand this in a more intuitive way, it is possible to refer to an example. If, on the one hand, the expected value of the operator $e^{i\xi\hat{q}+i\eta\hat{p}}$ is calculated:
\begin{equation}
Tr\left[\hat{\rho}e^{i\xi\hat{q}+i\eta\hat{p}}\right]=\iint dqdp\ e^{i\xi{q}+i\eta{p}}F(q,p,t),
        \label{eq:FdistrPS}   
\end{equation}
and on the other hand, the same is done for the operator
$e^{i\xi\hat{q}}e^{i\eta\hat{p}}$:
\begin{equation}
Tr\left[\hat{\rho}e^{i\xi\hat{q}}e^{i\eta\hat{p}}\right]=\iint dqdp\ e^{i\xi{q}+i\eta{p}}F(q,p,t),
        \label{eq:FdistrPS2}   
\end{equation}
 it is possible to verify that the function $F$ of Equation \eqref{eq:FdistrPS} is necessarily different from the function F of the equation \eqref{eq:FdistrPS2}. Consequently, it is possible to check that depending on the rule of association between operators and variables that is chosen, the pseudo-probability function will be different. Nevertheless, it is possible, according to Cohen \cite{COHEN96}, to define a general class of pseudo-probability functions in the following way:
\begin{equation}
Tr\left[\hat{\rho}f(\xi,\eta)e^{i\xi\hat{q}+i\eta\hat{p}}\right]=\iint dqdp\ e^{i\xi{q}+i\eta{p}}F(q,p,t),
        \label{eq:FdistrPS3}   
\end{equation}
or:
\begin{equation}
F^f=\frac{1}{4\pi^2}\iint d\xi d\eta Tr\{{\hat{\rho}f(\xi,\eta)e^{i\xi\hat{q}+i\eta\hat{p}}}\}e^{-i\xi q-i\eta p},
\label{eq:Fpseudo}
\end{equation}
A different choice in $f(\xi,\eta)$ corresponds to a different association rule, and therefore to a different function. Moreover, it is possible to use this equation to obtain the pseudo-probability function of a pure state
$\rho=|\psi\rangle\langle\psi|$:
\begin{multline}
F^f(q,p,t)=\frac{1}{4\pi^2}\int d\xi d\eta dq' \psi^*( q'+\frac{\hbar}{2}\eta,t) \\ \psi( q'-\frac{\hbar}{2}\eta,t) e^{-i\eta p}
 f(\xi,\eta)e^{i\eta(q'-q)}e^{-i\eta p}.
        \label{eq:F_purestate}
\end{multline}
In this way, it is clear that depending on the association rule chosen between the variables and operators, different pseudo-probability functions are obtained. The most commonly used ones in physics are listed in Table \ref{tabla1}.
\vspace{0.4cm}
\begin{table}[htbp]
\centering
\resizebox{8cm}{!}
{
\begin{tabular}{|c|c|c|} 
   \hline  
   Function & Rule & $f$  \\ \hline
   Wigner & $e^{i\xi{q}+i\eta{p}}
   \leftrightarrow
   e^{i\xi\hat{q}+i\eta\hat{p}}$ & 1  \\
   Standard-ordered & $e^{i\xi{q}+i\eta{p}}
   \leftrightarrow
   e^{i\xi\hat{q}}{e^{i\eta\hat{p}}}$ & $e^{-i\hbar\xi\eta/2}$ \\
    Kirkwood & 
    $e^{i\xi{q}+i\eta{p}}
    \leftrightarrow 
    e^{{i\eta\hat{p}}}e^{i\xi\hat{q}}$ &
    $e^{-i\hbar\xi\eta/2}$ 
     \\
      Husimi &$e^{i\xi{q}+i\eta{p}}=e^{v\beta^{*}-v^{*}\beta}
    \leftrightarrow e^{-v^{*}\hat{b}}e^{-v\hat{b}^{\dagger}}$& $e^{-|v|^2/2}$ \\ \hline
   \end{tabular}} 
   
   \caption{Association rules for the most commonly used pseudo-probability functions \cite{LEE95}.}
   \label{tabla1}  
\end{table}

Depending on the definition of the function $f(\xi$,$\eta$) it is possible to obtain different pseudo-probability functions. However, this work is centered on the simplest choice, $f(\xi$,$\eta$)=1, which is equivalent to Weyl association rule \cite{WEYL31}: 
\begin{equation}
e^{i\xi{q}+i\eta{p}}
\leftrightarrow
e^{i\xi\hat{q}+i\eta\hat{p}}.
\end{equation}
With this choice, equation \eqref{eq:Fpseudo} can be rewritten as:
\begin{multline}
W(q,p,t)=\frac{1}{4\pi^2}\iint d\xi d\eta\ \\ Tr\left[\hat{\rho}(\hat{q},\hat{p},t)e^{i\xi\hat{q}+i\eta\hat{p}}\right]  e^{-i\xi q-i\eta p},
\end{multline}
being this the function introduced by Wigner \cite{Wigner32}. For pure states it's possible to write:
    \begin{equation}
        W(q,p,t)=\frac{1}{2\pi}\int d\eta\  \psi^*( q+\frac{\hbar}{2}\eta,t)\psi( q-\frac{\hbar}{2}\eta,t) e^{-i\eta p}.
        \label{eq:WF_purestate}
    \end{equation}
With this function it is possible to obtain the expectation value of an operator as:
    \begin{equation}
        \langle\hat{A}\rangle=\iint dqdp\ W(q,p,t){A}(q,p,t).
        \label{eq:Average}
    \end{equation}
However, to use this expression it is necessary to obtain the Weyl transform of the operator. The simplest case is that of an operator that depends only on $\hat{x}$: 
\begin{multline}
    \hat{A}(\hat{x})=\int dy \hspace{1mm} e^{\frac{-iyp}{\hbar}} \bra{x+\frac{y}{2}}\hat{A}\ket{x-\frac{y}{2}} \\ =\int dy \hspace{1mm} e^{-\frac{-iyp}{\hbar}}  A(x-\frac{y}{2})\delta(y)= A(x).
\end{multline}
That is to say, the Weyl transform for such an operator will have the same expression, only changing the operator 
 $\hat{x}$ for the variable $x$. For an operator that depends exclusively on $\hat{p}$ it is possible to perform the same calculation, since it os possible to define the Weyl transform in the momentum representation as well \cite{brogaard2015wigner}. Therefore, another advantage of this formalism over other phase space formulations is that the expression of the Hamiltonian operators of the type $\hat{H}= \hat{T}(\hat{p}) + \hat{U}(\hat{x})$ in this space is simply  $H= T(p) + U(x)$. Finally, the expression for a general operator of the Weyl transform is as follows:
    \begin{multline}
        \hat{A}=\Phi\left[A\right]= \frac{1}{(2\pi\hbar)^2}\iiiint d\xi d\eta dqdp\  \\ A(q,p)e^{i(\xi(\hat{q}-q)+\eta(\hat{p}-p))}.
        \label{eq:WT}
    \end{multline}
Moreover, since it is an injective function, it has an inverse, thus, the inverse Weyl transform is defined as:
    \begin{equation}
        A(q,p)=\Phi^{-1}\left[\hat{A}\right]=\int d\eta\ e^{-i\eta p}\langle q+\frac{\hbar}{2}\eta|\hat{A}|q-\frac{\hbar}{2}
        \eta\rangle.
        \label{eq:IWT}
    \end{equation}
Nevertheless, with this expression it is difficult to obtain the Weyl transform of operators that depend on $\hat{q}$ and $\hat{p}$, consequently, in practice it is easier to use the expression introduced in \cite{MCCOY}:
\begin{equation}
    q^np^m=\frac{1}{2^n}\sum_{l=0}^n\binom{n}{l}\hat{q}^{n-l}\hat{p}^m\hat{q}^l.
\end{equation}

Once the Wigner function and the Weyl transform have been introduced it remains to be explained if, for a given Hamiltonian it is possible to obtain Wigner's function without solving the Schrödinger equation first. This can be done, and it is possible to define the dynamics of the system via the Moyal equation, which is an extension of the Liouville theorem \cite{Zachos56}:
\begin{equation}
\frac{\partial f}{\partial t}=\frac{H\star f-f\star H}{i\hbar}=\{\{H,f\}\},
\label{eq:corchetedemoyal}
\end{equation}
where the star product has been introduced \cite{Groenewold1946}:
\begin{equation}
\star=e^{i\frac{\hbar}{2}\left(\overset{\leftarrow}{\partial_q}\overset{\rightarrow}{\partial_p}-\overset{\leftarrow}{\partial_p}\overset{\rightarrow}{\partial_q}\right)}.
\end{equation}
The right-hand side of equation \eqref{eq:corchetedemoyal} is called the Moyal Bracket, and the quantum commutator is its Weyl correspondent. Furthermore, through a series expansion in $\hbar$ the Poisson bracket corrected by terms $\mathcal{O}(\hbar)$ can be obtained. In practice, it is possible to evaluate the star product via the Bopp shifts:

\begin{equation}
A(q,p)\star B(q,p)=A(q+\frac{1}{2}i\hbar\overset{\rightarrow}{\partial_p},p-\frac{1}{2}i\hbar\overset{\rightarrow}{\partial_q})\cdot B(q,p).    \label{eq:MoyalBShift}
\end{equation}
 With this expression it is possible to relate the Moyal equation with the energy of the system, obtaining an eigenvalue equation similar to the Schrödinger. This equation can be solved to find the Wigner function:
\begin{multline}
H(x,p)\star W(x,p)=W(x,p) \star H(x,p) \\ =H(x+\frac{1}{2}i\hbar\overset{\rightarrow}{\partial_p} , p-\frac{1}{2}i\hbar\overset{\rightarrow}{\partial_q})W(x,p)=EW(x,p).
\end{multline}

In the Wigner formalism there is  also a perturbation theory. One of the most important features of perturbation theory in the Wigner formalism is it being self-contained and therefore, it being formulated without reference to the conventional Hilbert space formulation \cite{Curtright_2001}.

Starting with a Hamiltonian decomposed into free and perturbed parts:
\begin{equation}
H=H_0 +\lambda H_1,
\end{equation}
where the eigenvalue equation is:
\begin{equation}
H(x,p)\star W(x,p)=W(x,p)\star H(x,p)=E_n(\lambda)W_n(x,p),
\end{equation}
it is possible to perform a series expansion of $E$ y $W$ in powers of $\lambda$ :
\begin{equation}
    E_n=E_n^0+\lambda E_n^1+\lambda^2E_n^2+\dots  
    \label{eq:potenciasenergía}
\end{equation} 
\begin{equation}
    W_n=W_n^0+\lambda W^1_n+\lambda^2W_n^2+\dots 
    \label{eq:potenciaswigner}
\end{equation}
 The equations for the first two orders, which are of particular interest for this work, are as follows:
\begin{equation}
    H_0\star W_n^0=E_n^0W_n^0,
    \label{eq:pert1}
\end{equation}
\begin{equation}
    H_0\star W_n^1+H_1\star W_n^0=E_n^0W_n^1+E_n^1W_n^0.
    \label{eq:pert2}
\end{equation}
Now, multiplying \eqref{eq:pert2} by $W_n^0\star$ from the left 
\begin{multline}
   W_n^0\star H_0\star W_n^1 + W_n^0\star H_1 \star W_n^0=E_n^1 W_n^0 \star  W_n^0 W_n^0 \\ + E_n^0 W_n^0\star W_n^0,
    \label{eq:pert3}
\end{multline}
and symplifying it using \eqref{eq:pert1}:
\begin{equation}
     W_n^0\star H_1 \star W_n^0= E_n^1 W_n^0\star W_n^0.
     \label{f_n}
\end{equation}
However, to solve this equation it is necessary to first introduce some properties of the star product and the Wigner function \cite{Curtright_2001}:
\begin{equation}
    W_{m} \star W_{n}=\frac{1}{2\pi\hbar}\delta_{nm}W_{n},
    \label{eq:pert4}
\end{equation}
\begin{equation}
    \int dx dp W_{m}(x,p)W_{n}^*(x,p)=\frac{1}{2\pi\hbar}\delta_{mn}.
    \label{eq:pert5}
\end{equation}
The relationship between two functions $f$ y $g$ of the phase space is also needed:
\begin{equation}
\int dxdp f \star g=\int dx dp \hspace{1mm}g\star f=\int dx dp \ f g.
\label{eq:pert6}
\end{equation}
Therefore, the equation \eqref{f_n} can be written as:
\begin{equation}
    \int dxdp \ E_n^1W_n^0\star W_n^0=\int dxdp \ W_n^0\star H_1\star W_n^0,
\end{equation}
it is possible to use equations \eqref{eq:pert4}, \eqref{eq:pert5} and \eqref{eq:pert6} to obtain the first-order energy correction:
\begin{equation}
    E_n^1=\int dxdp H_1 W_n^0.
    \label{eq:correnergía}
\end{equation}
\section{Black holes and the Schwarzschild metric}
\label{Schwarschild}
Once the Wigner-Weyl formalism has been derived, it is necessary to introduce the metric to be used to describe the black hole, the Schwarzschild metric.
The Schwarzschild metric was one of the first solutions ever found to the Einstein equations \cite{schwarzschild1999gravitational}. It does not only describe the gravitational field of an spherical star, but also allows us to describe static black holes, that is, those that do not have electric charge nor angular momentum. In general, the only parameter that will distinguish this type of black holes will be the mass. This geometry can be described by the line element:
\begin{multline}
    ds^2=-\left(1-\frac{2GM}{c^2r}\right)(cdt)^2+\left(1-\frac{2GM}{c^2r}\right)^{-1}(dr)^2 \\ +r^2(d\theta^2+sin^2\theta d\phi^2),
\end{multline}
And has the following properties \cite{hartle2003}:
\begin{itemize}
\item It is time independent, therefore, there is a Killing vector associated with this symmetry under displacements in the coordinate $t$. 
\item It has spherical symmetry. Thus there are also Killing vectors associated to with this symmetry. The most important one it is the associated with the  coordinate  $\phi$. The coordinate $r$ has a very direct interpretation arising from this symmetry. It is related to the area of the two-dimensional sphere of fixed $r$ and $t$.
\item In $r=2GM/c^2$ exists a coordinate singularity that can be solved, for instance, by using Eddington-Finkelstein coordinates. 
\end{itemize}

The laws of conservation of energy and angular momentum hold because the metric is independent of $t$ and $\phi$. With the aid of these laws it is  possible to derive an effective potential \cite{hartle2003}. In addition, during this calculation, natural units will be used, which will be reintroduced at the end of it. The explicit expressions of the conserved quantities are:
\begin{equation}
 e=-\xi\cdot u=\left(1-\frac{2M}{r}\right)\frac{dt}{d\tau},
\label{eq:energiaconservada}
\end{equation}
\begin{equation} 
 l=\eta \cdot u=r^2sin^2(\theta)\frac{d\theta}{d\tau},
\label{eq:momentoangular}
\end{equation}
where $e$ and $l$ are the energy and the angular momentum per unit rest mass, respectively, and where $u$ is the four-velocity. The angular momentum conservation implies that the orbit lies in a plane, as do the orbits in Newtonian theory. Therefore, it is possible to consider a specific angle $\theta$ without changing the physics. We have chosen  $\theta=\pi/2$ and $u^{\theta}=0$.
Finally, another conserved quantity can be deduced from the normalization of the four-velocity:
\begin{equation}
u\cdot u=g_{\alpha\beta}u^{\alpha}u^{\beta}=-1.
\label{eq:cuadri}
\end{equation}
Writing out \eqref{eq:cuadri} for the case of the Schwarzschild metric with the anterior condition: 
\begin{multline}
-\left(1-\frac{2M}{r}\right)(u^{t})^2+\left(1-\frac{2M}{r}\right)^{-1}(u^r)^2+r^2(u^{\phi})^2\\=-1.
\label{eq:sch1}
\end{multline}
Using \eqref{eq:energiaconservada} and \eqref{eq:momentoangular} can be rewritten as:
\begin{equation}
    \frac{e-1}{2}=\frac{1}{2}\left(\frac{dr}{d\tau}\right)^2+\frac{1}{2}\left( \left( 1-\frac{2M}{r}\right)\left(1+\frac{l^2}{r^2}\right)-1\right).
\end{equation}
Defining then the constant  $E= \frac{e^2-1}{2}$ and the potential $V_{\text{eff}}$ as:
\begin{equation}
    V_{\text{eff}}=-\frac{M}{r}+\frac{l^2}{2r^2}-\frac{Ml^2}{r^3},
\end{equation}
and reintroducing the constants:
\begin{equation}
    V_{\text{eff}}=\frac{1}{c^2}\left(-\frac{GM}{r}+\frac{l^2}{2r^2}-\frac{GMl^2}{c^2r^3}\right),
\end{equation}
it is possible to define and energy $E_{N}=mc^2E$ in order to have an analogous equation to the Newtonian case. Introducing, $L=l\cdot m$, the radial equation can be written as:
\begin{equation}
E_N=\frac{m}{2}\left(\frac{dr}{d\tau}\right)^2+\left(-\frac{GMm}{r}+\frac{L^2}{2mr^2}-\frac{GMl^2}{c^2mr^3}\right).    
\end{equation}
The final result is:
\begin{equation}
    V_{\text{eff}}=-\frac{GMm}{r}+\frac{L^2}{2mr^2}-\frac{GML^2}{c^2mr^3}.
    \label{eq:Veffnewt}
\end{equation}
This being the potential used in the next sections to solve the Schrödinger equation of a particle moving in this metric.

\section{Wave function for the Schwarzschild geometry}
\label{sec:Funciondeonda}
In this section, we calculate the wave function of an electron under the action of the Schwarzschild potential to obtain the Winger function later.
First, in analogy with the hydrogen atom, we define the gravitational 'Bohr radius':
\begin{equation}
    a_g=\frac{\hbar^2}{G\cdot m_e^2 \cdot M}=\frac{\hbar^2}{\kappa m_e}.
\end{equation}
Now, to make sense of the perturbative treatment, we calculate the order of magnitude of the mass from which it is possible to treat the relativistic term as a correction. This condition can be expressed as:
\begin{equation}
    \frac{GMm_e}{\braket{r}}\gg \frac{GM\braket{L^2}}{c^2m\braket{r^3}},
\end{equation}
This estimation has been done for the first excited state  $n=2$ and $l=1$, as this is the state for which the correction is calculated:
\begin{equation}
    \frac{\braket{r^3}}{\braket{r}} \gg \frac{2\hbar^2}{c^2m_e^2}.
    \label{condición}
\end{equation}
Using the values of the Table \ref{tabla2}:
\begin{equation}
    \frac{210a_g^3}{5a_g} \gg \frac{2\hbar^2}{c^2m_e^2}.
\end{equation}
As a consequence the mass has a superior limit:
\begin{equation}
    M \ll\frac{\sqrt{24}c\hbar}{Gm_e} \rightarrow M\ll1.6 \cdot 10^{16} \ \text{kg}.
\end{equation}
For the next sections the calculations have been done for a mass of $M= 10^{14} \ \text{kg}$, but it is possible to generalize for any value $M'$ that satisfies the condition for the upper limit. Using again the values of the Table \ref{tabla2} the radius will then be:
\begin{equation}
    \braket{r}= 3.97 \cdot 10^{-10}\cdot\left( \frac{10^{14} \ kg}{M'}\right) \ \text{m}.
\end{equation}
Proving as well that the Schwarzschild radius of this mass is less than the expected value of the position:
\begin{equation}
    r_s=\frac{2GM}{c^2}=1.48 \cdot 10^{-13} \cdot \left(\frac{M'}{10^{14} \ kg}\right) \ \text{m}.
\end{equation}

The mass obtained is too small, therefore the calculation won't be useful for stellar black holes. However, according to \cite{Montero_Camacho_2019} primordial black holes are in the correct mass range:
\begin{equation}
    6.65 \cdot 10^{13} \hspace{0.2 cm} \text{kg} <m<7.6 \cdot 10^{18} \hspace{0.2 cm} \text{kg}.
\end{equation}
Thus, the correction will be valid in this case. The calculation is then performed for the case of a bounded state of an electron and a primordial black hole in the range  $6.65 \cdot 10^{13} \ \text{kg}< M'\ll 1.6 \cdot 10^{16} \ \text{kg}$.

To obtain the electron's wave function we start from the Schwarzschild effective potential described by equation 
 \eqref{eq:Veffnewt}, and neglecting the relativistic correction we obtain the Newtonian potential, which is used to solve the Schrödinger equation:
\begin{equation}
    [\frac{\hbar^2}{2m_e}\nabla^2-\frac{Gm_eM}{r}-E]\cdot\psi=0.
    \label{eq:schrodinguer}
\end{equation}
Multiplying this equation by $4r$ from the left and using the Kustaanheimo-Stiefel transformation, which will be explained in more detail in \ref{ANEXO1}, Equation \eqref{eq:schrodinguer} can be written as \cite{4D}:
\begin{equation}
[\frac{\hbar^2}{2m}\nabla_4^2-4Es^2-4GMm_e]\cdot \psi=0,
\label{eq:osc}
\end{equation}
with:
\begin{equation}
\nabla_4^2=
\sum_{j=1}^{4}\frac{\partial^2}{\partial \xi_j^2}.
\end{equation}
With this equation an equivalence can be made between -4$E$ and $\frac{m_ew^2}{2}$, 4$GMm$ and  $N\hbar w$, where $N=n_1+n_2+n_3+n_4+2$, hence:
\begin{equation}
    E_N=\frac{-2G^2M^2m_e^3}{N^2\hbar^2}.
\end{equation}
On the other hand, the Schrödinger equation can be solved in a similar way to how it is solved for the hydrogen atom \cite{cuántica1}. This yields the energy spectrum:  $E_n=\frac{-G^2M^2m_e^3}{2n^2\hbar^2}$. Thus, the energy spectrum of this oscillator is equivalent to the energy of the Newtonian potential with $2n=N$.
By solving \eqref{eq:osc}, the next wave functions are obtained:
\begin{equation}
    \psi_{n_1n_2n_3n_4}=
\prod_{j=1}^{4}c_{n_j}\cdot e^{-(\alpha^2/2)\cdot \xi_j^2} H_{n_j}(\alpha\xi_j),
\label{eq:funciondeondaG}
\end{equation}
where $c_n=(\alpha/\sqrt{\pi}2^n n!)^{1/2}$ is a normalization constant, $H_{n}(\alpha\xi)$ the Hermite polynomials and $\alpha=(m_e\\w/\hbar)^{1/2}=({4}/{a_gN})^{1/2}$. 
However, in \cite{4D}, it was not considered that these waves functions do not have the correct degeneration. This is due to the fact that the constrain $p_w=0$ was not use in the momentum transformation \ref{eq:transmomentos}. Therefore, for the wave functions to be the correct ones, the constrain must be taken into account explicitly, which using \ref{eq:transmomentos} with $p_w=0$ can be written as:
\begin{equation}
\label{mañana2}
    L_{12}=\xi_1P_2-\xi_2P_1=\xi_3P_4-\xi_4P_3=L_{34}.
\end{equation}
This relation can be exploited using the phase space formalism, and as it will be proven later in Section \ref{L3} it is equivalent to $m_1=m_2$.
This condition implies that the hydrogen atom is equivalent to two isotropic harmonic oscillators with the same angular momentum $m$. With this condition the degeneration is correct, with $n^2$ states for each energy. However, since the quantum number $m$ is not well-defined on this basis, it will be difficult to determine which states are physical. To find them, one would need to express the wave functions $\psi_{nlm}$ as a linear combination of the wave functions  of Equation \eqref{eq:funciondeondaG}. This procedure has been generalized by Chen \cite{pepinillos}. As an example wi write down the first two states:
\begin{equation}
    \psi_{1,0,0} \propto \psi_{0,0,0,0}\,,
\end{equation}
\begin{equation}
\psi_{2,1,\pm1} \propto \left( \psi_{1,0,1,0}-\psi_{0,1,0,1}\pm i \psi_{0,1,1,0} \pm i \psi_{1,0,0,1}\right)\,.
\label{Wignerenergía}
\end{equation}

  It can be proven that in case of not using the constraint, on the one hand, there would be states with energies different from those of the hydrogen atom, which are the solutions with odd $N$. On the other hand, the degeneracy of the states with the correct energy correspond to that of a four-dimensional harmonic oscillator $\frac{(n-3)!}{n!3!}$ \footnote{
This formula is derived from the general case of the number of ways to distribute n indistinguishable objects into k boxes $\binom{n+k-1}{k-1}$.}.

\section{Wigner function for the Newtonian potential}
\label{sec:Wignerenergía}

Once the wave functions are obtained it is possible to obtain their Wigner functions like \cite{atomoH}:
\begin{multline}
W(\xi_1,\dots,\xi_4,p_1,\dots,p_4) = \\   
\frac{1}{\pi^4\hbar^4}
\int_{-\infty}^{\infty} \dots \int_{-\infty}^{\infty} dy_1\dots dy_4
e^{2i(q_1y_1+\dots+q_4y_4)/\hbar}\\
\psi^*(\xi_1+y_1,\dots,\xi_4+y_4)\psi(\xi_1-y_1,\dots,\xi_4-y_4),
\end{multline}
where $p_1,\dots, p_4$ are the conjugate momenta of  $\xi_1,\dots,\\\xi_4$, which satisfy the canonical commutation relations: $\{\xi_j,p_k\}$ $=\delta_{jk}$. Substituting \eqref{eq:funciondeondaG}:
\begin{multline}
\label{hoy}W(\xi_1,\dots,\xi_4,p_1,\dots,p_4)= \\ \frac{1}{\pi^4\hbar^4}
\int_{-\infty}^{\infty} \dots \int_{-\infty}^{\infty} dy_1\dots dy_4 
\prod_{j=1}^{4}|c_j|^2\cdot e^{-\alpha^2\cdot (\xi_j^2-y_j^2)}\\ e^{2iq_jy_j/\hbar}H_{nj}^*(\alpha(\xi_j+y_j))H_{nj}(\alpha(\xi_j-y_j)).
\end{multline}
Now, this expression must be worked on. First, consider the Hermite polynomials, and after adding and subtracting $\beta=iq/\alpha\hbar$, defining $z_j=\alpha(y-iq/\alpha^2\hbar^2)$ and using $H_n(-x)=(-1)^n H_n(x)$, it can be rewritten as:
\begin{multline}
   H_{nj}^*(\alpha(\xi_j+y_j))H_{nj}(\alpha(\xi_j-y_j))= \\ (-1)^n H_{nj}^*(z_j+\beta_j+\alpha\xi_j)H_{nj}((z_j-\beta_j-\alpha\xi_j).
\end{multline}
On the other hand, if Equation \eqref{hoy} is multiplied by $e^{\beta_j^2}e^{-\beta_j^2}$, simplified, and using the integration variable $z$ defined previously, the Wigner function becomes: 
\begin{multline}
W(\xi_1,\dots,\xi_4,p_1,\dots,p_4)= \\  
\frac{1}{\pi^6\hbar^4}\prod_{j=1}^{4} \frac{(-1)^{nj}}{2^{n_j}{n_j!}}  e^{-\alpha^2\xi_j^2+\beta_j^2}
\int_{-\infty}^{\infty} dz_j  e^{-z_j^2} \\ H_{nj}^*(z_j+\beta_j+\alpha\xi_j)H_{nj}(z_j-\beta_j-\alpha\xi_j),
\end{multline}
that using the expression defined in \cite{atomoH}:
\begin{multline}
    \int_{-\infty}^{\infty} dz e^{-z^2} H_{n}^*(z+\beta+\alpha\xi)H_{n}(z-\beta-\alpha\xi)= \\ \sqrt{\pi}2^nn!L_n(2(\alpha^2\xi^2-\beta^2)) ,
\end{multline}
 allows us to write the Wigner function as:
\begin{multline}
\label{mañana}
 W_{n_1,n_2,n_3,n_4}(\rho_1,\rho_2,\rho_3,\rho_4)= \\\frac{1}{\pi^4\hbar^4}\prod_{j=1}^{4}(-1)^{n_j}e^{-p_j^2/2}L_{n_j}(\rho_j^2),
\end{multline}
where we have defined $\rho=[2(\alpha^2\xi^2+q^2/(\alpha^2\hbar^2)]^{1/2}$. Additionally, as in the case of wave functions, the Wigner functions have the same constraint  $m_1=m_2$.
The ground state is then:
\begin{equation}
    W_{0,0,0,0}=\frac{1}{\pi^4\hbar^4}e^{-(\rho_1^2+\rho_2^2+\rho_3^2+\rho_4^4)/2}.
    \label{fundamental}
\end{equation}
Meanwhile, for the first excited state $\psi_{2,1,1}$,  it is possible to define the Wigner function from 
Equation \eqref{Wignerenergía}:
\begin{equation}
    W_{2,1,1} \propto \left( W_{1,0,1,0}-W_{0,1,0,1} + i W_{0,1,1,0} + i W_{1,0,0,1}\right),
\end{equation}
that using the values of the Equation \eqref{mañana} it is equal to:
\begin{multline}
    W_{2,1,1}=\frac{-i}{2\pi^4\hbar^4}e^{-(\rho_1^2+\rho_2^2+\rho_3^2+\rho_4^4)/2} ( L_{1}(\rho_1^2)L_{1}(\rho_3^2) \\ - L_{1}(\rho_2^2)L_{1}(\rho_4^2) +iL_{1}(\rho_2^2)L_{1}(\rho_3^2)+iL_{1}(\rho_1^2)L_{1}(\rho_4^2) ),
\end{multline}
with the Laguerre polynomial for $n=1$:
\begin{equation}
    L_1(x)=-x+1.
\end{equation}

\label{L3}

Finally, it is possible to calculate the Wigner function in the basis of $L_3$ and $H$. With this calculation the constraint condition is justified. Moreover, these Wigner functions explicitly satisfy this condition and are therefore valid Wigner functions. To find them, we start from the eigenvalue equation in phase space \cite{CAMPOS201860}:
\begin{equation}
H(x,p)\star W(x,p)=E\cdot W(x,p),
\label{eq:autovalores}
\end{equation}
where the general Hamiltonian of the problem is:
\begin{equation}
    H=\frac{p_x^2+p_y^2+p_z^2}{2m_e}-\frac{\kappa}{r}-\frac{\lambda}{r^3}.
    \label{eq:Hamiltoniano}
\end{equation}
with $\kappa=GMm_e$ and $\lambda=\frac{GML^2}{c^2mr^3}$. To calculate the solutions of \eqref{eq:autovalores} for this Hamiltonian is too difficult. Therefore the same procedure as in Section \ref{sec:Funciondeonda} will be carried out, transforming the coordinates of the problem:
\begin{equation}
    H=\frac{1}{8rm_e}\sum_{i=1}^4P_i^2-\frac{\kappa}{r}-\frac{\lambda}{r^3}.
\end{equation}
But, again it should be taken into account the restriction $p_w=0$, described by equation \eqref{mañana2}.
Additionally, using the hipersurface of phase space $H=E$ it can be written:
\begin{equation}
     \frac{1}{2m_e}\sum_{i=1}^4P_i^2-4{\kappa}+4\frac{\lambda}{r^2}=4rE,
\end{equation}
which, when the perturbative term is neglected, becomes:
 \begin{equation}
     \frac{1}{2m_e}\sum_{i=1}^4P_i^2-4{\kappa}=4s^2E.
\end{equation}
This equation can be solved using the same procedure that in section \ref{sec:Funciondeonda}, making the equivalence between -$4E$ and $\frac{m_ew^2}{2}$, 4$\kappa$ and  $N\hbar w$, where $N=n_1+n_2+n_3+n_4+2$, and working with the effective Hamiltonian of an harmonic oscillator in four dimensions:
\begin{equation}
    H=\frac{1}{2m_e}\sum_{i=1}^4P_i^2+s^2(\frac{m_e \omega^2}{2}).
\end{equation}
Therefore, the eigenvalue equation simplifies to:
\begin{equation}
\label{618}
    \left(\frac{1}{2m_e}\sum_{i=1}^4P_i^2+s^2\frac{m_e \omega^2}{2}\right) \star W(x,p)=(\hbar \omega N)W(x,p).
\end{equation}
Moreover, the angular momentum constraint implies one more equation:
\begin{equation}
\label{619}
   (L_{12} \pm L_{34})\star W=0.
\end{equation}
From this equation, the constraint condition in terms of the quantum numbers  $m$  can already be deduced, it is equivalent to:
\begin{equation}
\label{memento}
    m_{12}=m_{34},
\end{equation}
where according to Bayen, Flato and Fronsdal \cite{BAYEN1978111}, $m$ is a quantum number that in the case of a bidimensional harmonic oscillator can take the following values: $m_{ij}=n_{i}+n_{j}, n_i + n_j-2, \dots, -n_i-n_j$. Therefore, the previous condition can be rewritten in terms of the quantum numbers  $n_1,n_2,n_3,n_4$ \cite{CAMPOS201860} as:
\begin{equation}
    n_1+n_2=n_3+n_4.
  \label{eq:n1n2}
\end{equation}
Furthermore, one can solve the system of equations formed by  \eqref{618} and \eqref{619} in this basis assuming that the Wigner function is separable:
\begin{equation}
    W=f_{12}(\xi_1,\xi_2,p_1,p_2)\cdot f_{34}(\xi_3,\xi_4,p_3,p_4).
\end{equation}
It leaves two systems of equations:
\begin{equation}
    \frac{1}{2m_e}\sum_{j=1}^2P_i+\sum_{j=1}^2\xi_j^2 \frac{m_e \omega^2}{2}\star f_{12}=C_{12}f,
\end{equation}
\begin{equation}
    L_{12}\star f=M_{12}\cdot f.
\end{equation}
And:
\begin{equation}
    \frac{1}{2m_e}\sum_{j=3}^4P_i+\sum_{j=3}^4\xi_j^2 \frac{m_e \omega^2}{2}\star f_{34}=C_{34}f,
\end{equation}
\begin{equation}
    L_{34}\star f=M_{34}\cdot f,
\end{equation}
where $C_{12}+C_{34}=N\hbar w$. These equations are completely analogous and can be written as follows:
\begin{equation}
    \left(H_{ij}\pm L_{ij}-\left(\frac{C_{ij}}{\omega}\pm M_{ij}\right)\right)\star f_{ij}=0,
\end{equation}
where the Hamiltonian using the nomenclature of the section  \ref{sec:Wignerenergía}, can be written as:
\begin{equation}
    H_{ij}=\frac{1}{2}\left((\alpha^2\hbar)^{-1}\sum_{k=i}^jP_k^2+\alpha^2\hbar\sum_{k=i}^j\xi_k^2)\right).
\end{equation}

To solve this equation it is necessary to use the Bopp shifts $\xi \rightarrow \xi + \frac{i\hbar}{2}\partial p$ y $p \rightarrow p - \frac{i\hbar}{2}\partial\xi$.  This will give two equations, one for the real part and one for the complex part. The second can be written as:
\begin{multline}
(\frac{\pm\partial\xi_i}{\sqrt{\omega m_e}}+\partial p_j\sqrt{\omega m_e})(\sqrt{\omega m_e}\xi_j\mp\frac{p_i}{\sqrt{\omega m_e}}) \\-(\frac{\partial\xi_j}{\sqrt{\omega m_e}}\mp\partial p_i\sqrt{\omega m_e})(\sqrt{\omega m_e}\xi_i\pm \frac{p_j}{\sqrt{\omega m_e}})f_{ij}=0,
\end{multline}
and thus restricts the function $f_{ij}$  to depend on a single variable $z=\frac{2}{\hbar}(H_{ij}\pm L_{ij})$.
Therefore, the real part is reduced to an ordinary differential equation:
\begin{equation}
    \left(\frac{1}{4}z-z\partial_z^2-\partial_z-\frac{1}{2\hbar}\left(\frac{C_{12}}{\omega}\pm M_{ij} \right)\right)f(z)=0,
\end{equation}
using $f(z)=e^{-z/2}L(z)$ one can obtain the Laguerre equation:
\begin{equation}
    \left(z\partial_z^2+(1-z)\partial_z+\left(\frac{C_{12}}{2\omega\hbar}\pm \frac{M_{ij}}{2\hbar}-\frac{1}{2}\right)\right)L(z)=0.
\end{equation}
where $C_{ij}$ is the constant of a two-dimensional harmonic oscillator, and $M_{ij}=\hbar\cdot m_{ij}$. The solution to this equation is given by Laguerre polynomials $L_n(z)$ with $n$=${(n_{ij}\pm m_{ij})}/{2}$.
Thus, once this calculation is done, the following Wigner functions are obtained:
\begin{multline}
    W_{n_{12},n_{34},m_{12},m_{34}}=\frac{1}{\pi^4\hbar^4}e^{-\frac{2}{\hbar} \left( H_{12}+H_{34} \right)}\\ L_{(n_{12}+m_{12})/2}\left(\frac{2}{\hbar}(H_{12}+L_{12})\right) \\ L_{(n_{12}-m_{12})/2}\left(\frac{2}{\hbar}(H_{12}-L_{12})\right) \\  L_{(n{34}+m_{34})/2}\left(\frac{2}{\hbar}(H_{34}+L_{34})\right) \\ L_{(n34-m_{34})/2}\left(\frac{2}{\hbar}(H_{34}-L_{34})\right),
\end{multline}
with $m_{ij}=n_{ij},n_{ij}-2,\dots,-n_{ij}$, and $n_{12}=n_{34}$.
 The Wigner function of the ground state is:
   \begin{equation}
    W_{0,0,0,0}=\frac{1}{\pi^4\hbar^4}e^{-\frac{2}{\hbar} (H_{12}+H_{34})}
    \label{wig0},
\end{equation}
that using $\rho=[2(\alpha^2\xi^2+q^2/(\alpha^2\hbar^2)]^{1/2}$, it is the same that the one obtained in the Section \ref{sec:Wignerenergía} as expected:

\begin{equation}
   W_{0,0,0,0}=\frac{1}{\pi^4\hbar^4}e^{-(\rho_1^2+\rho_2^2+\rho_3^2+\rho_4^4)/2}.
\end{equation}

Additionally, there will again be four excited states: $W_{1,1,1,1}$, $W_{1,1,-1,-1},W_{1,1,1,-1}$ and $W_{1,1,-1,1}$. Writing one of them explicitly:
\begin{multline}
W_{1,1,1,1}=\frac{1}{\pi^4\hbar^4}[1-\frac{1}{\alpha^2\hbar^2}(P_1^2+P_2^2) \\-{\alpha^2}(\xi_1^2+\xi_2^2) +\frac{2}{\hbar}(\xi_1P_2-\xi_2P_1)]\\
\times [1-\frac{1}{\alpha^2\hbar^2}(P_3^2+P_4^2)-{\alpha^2}(\xi_3^2+\xi_4^2)+\frac{2}{\hbar}(\xi_3P_4-\xi_4P_3)] \\
\times e^{-(\rho_1^2+\rho_2^2+\rho_3^2+\rho_4^4)/2}.
\end{multline}

\section{Explicit computations through the harmonic duality}
\label{energía}
In this section, we have calculated some expected values for the ground state and the first excited state. To achieve this, as a result of multiplication by 4r in the Schrödinger equation, the Wigner function must be normalized to, $dV=s^5ds$ \cite{s5} so the correspondence is exact.

First, we  find the expected value of the position, the uncertainty, and the energy. The expected values can be compared with those in Table \ref{tabla2}, which proves that the same results are obtained as with the Hilbert space formalism.
Starting with the ground state:
\begin{equation}
    W_{0,0,0,0} \propto e^{-(\rho_1^2+\rho_2^2+\rho_3^2+\rho_4^4)/2},
    \label{wig0}
\end{equation}
and normalizing it:
    \begin{equation}
    W_{0,0,0,0} =\frac{8\alpha^2}{\hbar^4\pi^2} e^{-(\rho_1^2+\rho_2^2+\rho_3^2+\rho_4^4)/2},
    \label{wig1}
\end{equation}
it is possible to calculate the following expected values as:
\begin{multline}   \braket{r^n}=\int_{0}^{\infty} s^{5+2n} ds \int_{0}^{\pi/2} \frac{sen(2\delta)}{2} \int_{0}^{2\pi} d\beta \\ \int_{\beta}^{\beta-2\pi}d\gamma \int_{-\infty}^{\infty} dp_1 \dots dp_4 \hspace{0.3cm} {W_{0,0,0,0}} .
\end{multline}
As a result, the following has been obtained:
\begin{equation}
    \braket{r}\frac{3}{\alpha^2}=\frac{3a_g}{2},
\end{equation}
and:
\begin{equation}
    \braket{r^2}=\frac{12}{\alpha^4}=3a_g^2.
\end{equation}
The expected value of the position is then:
\begin{multline}
    \braket{r}=1.19 \cdot 10^{-10} \left(\frac{10^{14}}{M'}\right)\hspace{0.5mm} \text{m}, \\ \Delta r = 6.89  \cdot 10^{-11}\left(\frac{10^{14}}{M'}\right)^2\hspace{1mm} \text{m} ,
\end{multline}
where $\Delta r=\sqrt{\braket{r^2}-\braket{r}^2}$. Finally, for this state, the first-order correction to the energy has also been calculated. The correction due to general relativity for the ground state will be zero, since this state has $l=0$. However, a quantum correction to the Newtonian potential exists \cite{Donoghue_1994}, such that for the states with  $l\neq0$ will be smaller than the general relativity correction, but, in the case of the ground state it becomes the dominant correction. The correction is therefore:  
\begin{equation}
    \lambda H_1=\frac{127G^2Mm_e\hbar}{30\pi^2r^3c^3},
\end{equation}
and using this Hamiltonian in the expression \eqref{eq:correnergía} the correction is: 
\begin{multline}
    E_2^{(1)}=\int d\xi_1 \dots d\xi_4 \int dp_1 \dots dp_4  \\
    \frac{W_{0,0,0,0}(\xi_1,\dots,\xi_4,p_1,\dots,p_4)}{(\xi_1^2+\xi_2^2+\xi_3^2+\xi_4^2)^6}.
\end{multline}
This integral is divergent. Therefore, a lower limit  $\epsilon$ must be imposed on the variable $s$. Performing a change of variable again, the integral becomes:
\begin{multline}
    E_2^{(1)}=\frac{8\alpha^2}{\hbar^4\pi^2}\int_{\epsilon}^{\infty}ds \int_{0}^{\pi/2} \frac{sen(2\delta)}{2} \int_{0}^{2\pi} d\beta \\\int_{\beta}^{\beta-2\pi}d\gamma  \int_{-\infty}^{\infty} dp_1 \dots dp_4 \ \frac{{e^{-\alpha^2s^2}}}{s}  \times e^{\frac{-1}{\alpha^2\hbar^2}(p_1^2+p_2^2+p_3^2+p_4^2)} \\ =\frac{\alpha^6}{2}\Gamma[0,\alpha^2\epsilon]=\frac{4}{a_g^3}\Gamma[0,2\epsilon/a_g]. \end{multline}

The function $\Gamma[0,x]$ \footnote{This function is the incomplete gamma function defined as $\Gamma[a,x]=\int_x^{\infty} t^{a-1}e^{-t} dt$.}, with  $x$ approaching 0, is divergent, so we will study how it diverges. To observe how this function approaches zero, it is possible to perform a series expansion, obtaining:
\begin{equation}
    \Gamma[0,x]=-\text{log}(x)-\gamma + x + \frac{x^2}{2}+ \dots
\end{equation}
which allows us to see that the divergence for this correction is logarithmic. Thus, it is possible to express the first-order energy as:
\begin{equation}
E_2=-\frac{G^2M^2m^3}{2\hbar^2}+\frac{127G^2Mm_e\hbar}{30\pi^2c^3}\frac{1}{4a_g^3}\text{log}(2\epsilon/a_g).
\end{equation}

The calculations from the previous section  are repeated for the Wigner function $W_{2,1,1}$  obtained in section \ref{sec:Wignerenergía}. Again, this function has to be normalized with $s^5 ds$ to replicate the results of classical quantum theory. Therefore, starting with:
\begin{multline}
    W_{2,1,1} \propto e^{-(\rho_1^2+\rho_2^2+\rho_3^2+\rho_4^4)/2} (L_{1}(\rho_1^2)L_{1}(\rho_3^2) \\ - L_{1}(\rho_2^2)L_{1}(\rho_4^2) +iL_{1}(\rho_2^2)L_{1}(\rho_3^2)+iL_{1}(\rho_1^2)L_{1}(\rho_4^2)),
\end{multline}
with the normalization condition:
\begin{multline}
    1=C\int_{0}^{\infty} s^5 ds \int_{0}^{\pi/2} \frac{sen(2\delta)}{2} \int_{0}^{2\pi} d\beta \\ \int_{\beta}^{\beta-2\pi}d\gamma \int_{-\infty}^{\infty} dp_1 \dots dp_4 \hspace{0.3cm} W_{2,1,1},
\end{multline}
we obtain the following Wigner function:
\begin{multline}
    W_{2,1,1}=\frac{-i\alpha^2}{8\pi^4\hbar^4}(e^{-(\rho_1^2+\rho_2^2+\rho_3^2+\rho_4^4)/2}\left( L_{1}(\rho_1^2)L_{1}(\rho_3^2)\right) \\ - L_{1}(\rho_2^2)L_{1}(\rho_4^2)+iL_{1}(\rho_2^2)L_{1}(\rho_3^2)+iL_{1}(\rho_1^2)L_{1}(\rho_4^2)).
\end{multline}
Again, the position and the uncertainty are calculated. For this, the expected values of $r$ and $r^2$ are needed:
\begin{multline}
    \braket{r^n}=\int_{0}^{\infty} s^{5+2n} ds \int_{0}^{\pi/2} \frac{sen(2\delta)}{2} \int_{0}^{2\pi} d\beta \\ \int_{\beta}^{\beta-2\pi}d\gamma \int_{-\infty}^{\infty} dp_2 \dots dp_4\hspace{0.3cm} W_{2,1,1},
\end{multline}
resulting in:
\begin{equation}
    \braket{r}=5a_g= 3.97 \cdot 10^{-10}\left( \frac{10^{14} \ \text{kg}}{M'}\right) \ \text{m},
\end{equation}   
\begin{equation}
\Delta r
=\sqrt{5}a_g^2=1.78 \cdot 10^{-10}\left(\frac{10^{14} \text{kg}}{M'}\right) \ \text{m}.
\end{equation}
The energy correction for this state is also calculated:
\begin{multline}
     E_2^{(1)}=\int_{0}^{\infty}  ds \int_{0}^{\pi/2} \frac{sen(2\delta)}{2} \int_{0}^{2\pi} d\beta\int_{\beta}^{\beta-2\pi}d\gamma \\ \int_{-\infty}^{\infty} dp_1 \dots dp_4 \hspace{0.3cm} \frac{W_{2,1,1}}{s},
\end{multline}
obtaining the following result:
\begin{equation}
    E_2^{(1)}=\frac{\alpha^6}{24}=\frac{1}{24a_g^3}.
\end{equation}
In general, the Wigner functions obtained through the Kustaanheimo-Stiefel transformation do not have 
a well-defined quantum number $l$ well-defined. However, a wave function that does meet this requirement has been chosen, as the general relativity correction does depend on $l$. It is possible to verify, checking  Table \ref{tabla2} in Appendix \ref{ANEXO2}, that the value obtained for this correction is the same as that obtained by the Schrödinger formalism. This demonstrates that the Wigner function, as defined here, has a well-defined quantum number. It allows for an exact calculation of the general relativity correction, using $L^2=\hbar^2 l(l+1)$ with $l=1$:
\begin{equation}
    E_2=-\frac{G^2M^2m^3}{8\hbar^2}-\frac{GM2\hbar^2}{c^2m_er^3}\frac{1}{24a_g^3}.
\end{equation}
 Numerically the value of the first term of the energy is:
\begin{equation}
    E_2^{(0)}-\frac{G^2M^2m^3}{8\hbar^2}=-9.53\cdot 10^{-18}\left( \frac{M'}{10^{14} \ \text{kg}}\right)^2\text{J},
\end{equation}
and that of the first correction:
\begin{equation}
    E_2^{(1)}=-\frac{GM2\hbar^2}{c^2m_e}\frac{1}{24a_g^3}=-5.91\cdot 10^{-21} \left( \frac{M'}{10^{14} \ \text{kg}}\right)^4 \ \text{J}.
\end{equation}

In fact, by examining the expected values, it can be verified that they are identical to those of the radial function $R_{21}$. Therefore, it is evident that using the phase space formalism allows us to replicate the results of the usual perturbation theory. Consequently, it will be possible to extend this procedure to other excited states, provided that we establish the relationship between the wave functions in the oscillator basis and the wave functions in the  $n,l,m$ basis.

\section{Conclusions}
We have review the main tools of the phase space formalism were introduced in order to clarify the main properties of the Wigner approach. Although the Husimi function and the Segal-Bargmann space are interesting, we have focused on the Wigner function and the Weyl transform characterization. Using these tools, it was established a perturbation approach that allows to work in a symplectic formalism within a curved spacetime. Following this method, we have studied quantum dynamics of elementary particles on a Schwar-zschild geometry. This approach can be applied, for instance, to the bound state of an electron and a primordial black hole. Moreover, we have used an effective potential for this metric that can be derive using the laws of conservation of energy and angular momentum. Finally, the different approximations assumed in the computation have been discussed.\\

At the leading order, solving this equation is equivalent to solving for the Hydrogen atom; and using spherical coordinates it is possible to obtain the wave functions $\psi_{\text{nlm}}$. However, in this work  this approach was not used. Instead, the Kustaanheimo-Stiefel transformation was employed to connect the harmonic potential with a potential that depends on the inverse of the distance, such as the Coulomb or, in this specific case, the Newtonian potential. Once these wave functions were obtained, the calculations from Nouri article \cite{atomoH} were reproduced, achieving Wigner functions through the Weyl transform of the previous wave functions. We have updated these results by considering the momentum constrain in order to derive the correct Wigner functions.\\

Subsequently, the Wigner pseudo-probability distribution has been obtained in a different basis. For this purpose, the calculations of Campos, Martins and Fernandes \cite{CAMPOS201860} have been reproduced. This demonstrated, on the one hand, that, similar to the Hilbert space formalism, Wigner functions can be obtained on two different bases; and, on the other hand, that it is also possible to obtain Wigner functions without referencing the Schrödinger equation, thus verifying that the formalism is self-contained. Finally, in section \ref{energía}, we have calculated the position, uncertainty and the first-order correction to the energy of the first two excited states using the perturbation theory of the symplectic formalism. We have verified that these results can be reproduce by using the standard quantum theory in configuration space within the Hilbert space formalism, with results summarized in Table \ref{tabla2}. \\

\bigskip
\section*{Acknowledgements}

This work is partially supported by the project PID2022-139841NB-I00 funded by MICIU/AEI/10.13039/501100 011033 and by ERDF/EU, and by COST
(European Cooperation in Science and Technology) Actions CA21 106, CA21136, CA22113 and CA23130.

\appendix
\titleformat{\section}[block]{\normalfont\Large\bfseries}{Appendix \thesection.}{1em}{}
\addtocontents{toc}{\protect\setcounter{tocdepth}{0}}
\section{Kustaanheimo- \newline
Stiefel transformation}
\addtocontents{toc}{\protect\setcounter{tocdepth}{2}}
\addcontentsline{toc}{section}{Appendix \thesection. Kustaanheimo-Stiefel transformation}
\label{ANEXO1}
  The Kustaanheimo-Stiefel transformation has been used to solve Schröndinger equation, thus this appendix will be used to understand it in more detail. This is an application that can be expressed as:

\begin{equation} \label{Trans}
\begin{pmatrix}
x \\
y \\
z \\
0
\end{pmatrix}=2
\begin{pmatrix}
\xi_3 &- \xi_4 & \xi_1 & -\xi_2\\
\xi_4 & \xi_3 & \xi_2 & \xi_1\\
\xi_1 & \xi_2 & -\xi_3& -\xi_4 \\
\xi_2 & -\xi_1 & -\xi_4 & \xi_3
\end{pmatrix}
\begin{pmatrix}
\xi_1 \\
\xi_2 \\
\xi_3 \\
\xi_4
\end{pmatrix}.
\end{equation}
That is:
\begin{eqnarray} \label{coordenadas} \nonumber
x= 2(\xi_1\xi_3-\xi_2\xi_4), \\ \nonumber
    y= 2(\xi_1\xi_4+\xi_2\xi_3), \\ 
    z= \xi_1^2+\xi_2^2-\xi_3^2-\xi_4^4.
\end{eqnarray}
Another possible way to parameterize this transformation is:
\begin{eqnarray} \label{esfericas} \nonumber
\xi_1 =s\cdot cos(\delta)cos(\beta), \\  \nonumber
\xi_2=s\cdot cos(\delta)sin(\beta), \\ \nonumber
\xi_3=s\cdot sin(\delta)cos(\gamma),\\
\xi_4=s\cdot sin(\delta)sin(\gamma),\\ \nonumber
\end{eqnarray}
where:
\begin{equation}
    s=r^{1/2}, \hspace{0.5cm} 2\delta=\theta, \hspace{0.5cm} \beta \pm \gamma=\phi.
\end{equation}
With his parameterization, it is possible to understand the meaning of this transformation. This defines a surjective map, where if a point $\xi \in \mathbb{R}^{4*}$ has the image $q \in \mathbb{R}^{3*} $, then every point $\xi' \in \mathbb{R}^{4*}$ defined as:

\begin{equation}
\begin{pmatrix}
\xi_1' \\
\xi_2' \\
\xi_3' \\
\xi_4'
\end{pmatrix}=
\begin{pmatrix}
\cos(\phi) &\sin(\phi) & 0 & 0\\
-\sin(\phi) & \cos(\phi) & 0 & 0 \\
0 & 0 & \cos(\phi)& -\sin(\phi) \\
0 & 0 & \sin(\phi) & \cos(\phi)
\end{pmatrix}
\begin{pmatrix}
\xi_1 \\
\xi_2 \\
\xi_3\\
\xi_4
\end{pmatrix},
\label{eq:transaccion}
\end{equation}
has the same image $q \in \mathbb{R}^{3*} $. Moreover, the set \{$R_\phi: \phi \in (0,2\pi)\}$ generates a Lie group isomorphic to $SO(2) \times SO(2)$. In this way, the angular momentum associated with this rotation is the generator of a U(1) Lie group.
Using Equations\eqref{eq:transcoordenadas} and \eqref{esfericas} is possible to obtain the differential mapping:
\begin{equation}
\begin{pmatrix}
dx \\
dy \\
dz \\
dw
\end{pmatrix}
=
2\begin{pmatrix}
\xi_3 &- \xi_4 & \xi_1 & -\xi_2\\
\xi_4 & \xi_3 & \xi_2 & \xi_1\\
\xi_1 & \xi_2 & -\xi_3& -\xi_4 \\
\xi_2 & -\xi_1 & -\xi_4 & \xi_3
\end{pmatrix}
\begin{pmatrix}
\xi_1 \\
\xi_2 \\
\xi_3 \\
\xi_4
\end{pmatrix},
\end{equation}
where $w$ is a "dummy coordinate" such that $dw$ is not a total differential. From this equation is possible to derive the transformation of the linear momenta which is \cite{CAMPOS201860}:
\begin{equation}
\begin{pmatrix}
p_x \\
p_y \\
p_z \\
p_w
\end{pmatrix}=\frac{1}{2s^2}
\begin{pmatrix}
\xi_3 &- \xi_4 & \xi_1 & -\xi_2\\
\xi_4 & \xi_3 & \xi_2 & \xi_1\\
\xi_1 & \xi_2 & -\xi_3& -\xi_4 \\
\xi_2 & -\xi_1 & -\xi_4 & \xi_3
\end{pmatrix}
\begin{pmatrix}
P_1 \\
P_2 \\
P_3 \\
P_4
\end{pmatrix}.
\label{eq:transmomentos}
\end{equation}
For Equations \eqref{eq:transcoordenadas} and \eqref{eq:transmomentos}  to define a canonical transformation , the Poisson brackets must be independent of the chosen coordinates. If we evaluate them in the 4-coordinate system, we obtain:
\begin{eqnarray}
    \{q_i,q_j\}=0,\hspace{0.5 cm} \{q_i,p_j\}=\delta_{ij},\hspace{0.5 cm}\{q_i,p_w\}=0 ,\nonumber\\ 
\{p_i,p_j\}=\sum_{k=1}^3\epsilon_{ijk}\frac{q_k}{q^2}p_w,\hspace{0.5 cm} \{p_i,p_w\}=\frac{q_i}{q^2}p_w.
\end{eqnarray}
Thus, it is possible to verify that the transformation is canonical if and only if $p_w=0$.

\addtocontents{toc}{\protect\setcounter{tocdepth}{0}}
\section{Calculation in the Schrödinger formalism}
\addtocontents{toc}{\protect\setcounter{tocdepth}{2}}
\addcontentsline{toc}{section}{Appendix \thesection. Calculation in the Schrödinger formalism}\label{ANEXO2} The first-order correction to the energy in Hilbert space is:
\begin{equation}   E_n^{(1)}=\bra{\psi_{n}^{(0)}} H_1 \ket{\psi_{n}^{(0)}}.
\end{equation}
And the expected values are:
\begin{equation} \braket{r}=\bra{\psi_{n}^{(0)}} r^n \ket{\psi_{n}^{(0)}}.
\end{equation}
In Table \ref{tabla2} are listed the main expected values that are needed to prove that the results obtained in the symplectic formalism are equivalent to the ones obtained in Hilbert space. 
\begin{table}[htbp]

\centering
\resizebox{8cm}{!}{
\begin{tabular}{|c|c|c|c|c|c|}
   \hline
   State & Radial function & $\braket{\tilde{r}}$ &$\braket{\tilde{r}^2}$ &$\braket{\frac{1}{\tilde{r}^3}}$ & $\braket{\tilde{r}^3} $\\
   \hline
   $R_{10}$ & $2(\frac{1}{a_g})^{3/2}e^{{-\tilde{r}}}$& $\frac{3}{2}$ & $3$ &${4}\Gamma[0,\frac{2\epsilon}{a_g}]$&$\frac{15}{2}$ \\

   \hline
    $R_{21}$ & $\frac{1}{\sqrt{3}}(\frac{1}{2a_g})^{3/2}\tilde{r}\cdot e^{-\frac{\tilde{r}}{2}}$ & $5$ & $30$ &$\frac{1}{24}$ & $210$   \\
    \hline
   \end{tabular}}
   
   \caption{Wave functions and expected values in Hilbert space ($\tilde{r}=\frac{r}{a_g}$).}
   \label{tabla2}
   
\end{table}

\printbibliography

\end{document}